\documentstyle[epsf,twoside,fleqn,espcrc2]{article}

\newcommand{\bee}{\begin{equation}}
\newcommand{\ee}{\end{equation}}
\newcommand{\beea}{\begin{eqnarray}}
\newcommand{\eea}{\end{eqnarray}}

\newcommand{\ewxy}[2]{\setlength{\epsfxsize}{#2}\epsfbox[40 140 670 650]{#1}}

\newcommand{\AmS}{{\protect\the\textfont2
  A\kern-.1667em\lower.5ex\hbox{M}\kern-.125emS}}

\title{Instanton Content of the SU(3) Vacuum\thanks{Talk presented by
Chet Nieter}}

\author{Anna Hasenfratz\address{Department of Physics,
University of Colorado, Boulder, CO 80309-390} and Chet
Nieter$^\textrm{\scriptsize a}$}

\begin{document}

\begin{abstract}We study the topological content of the SU(3) 
vacuum using a method based on RG mapping developed for SU(2) gauge
theory earlier. RG mapping, in which a series of APE-smearing steps is
done while tracking the observables, reduces the short range
fluctuations in the gauge fields while preserving the long
structure. This allows us to study the instanton size distribution and
topological susceptibility for SU(3) gauge theory. We arrive at a value
for the topological susceptibility ${\chi}^{1/4}$, of 203(5) MeV. The
size distribution peaks at $\rho = 0.3fm$, and is in good agreement with
the prediction of the instanton liquid models.
\end{abstract}

% typeset front matter (including abstract)
\maketitle

Instantons play an essential role in the QCD vacuum.  They explain the
U(1) problem \cite{U1} and there is growing evidence that they are
responsible for chiral symmetry breaking and the low energy hadron
spectrum \cite{Negele,SU2_SPECTR}.  Phenomenological instanton liquid
models describe the propagation of quarks as hopping from instanton to
instanton.  This requires the instantons and anti-instantons to overlap
to provide continuous paths for this propagation.  To understand if
these paths are formed one has to determine the location and size
distribution of the instantons in the vacuum.

Because instantons carry only a few percent of the action, they are
hidden by vacuum fluctuations.  To identify the instantons some method
to reduce the short-range quantum fluctuations in the gauge fields while
preserving the topological content of the vacuum is needed.  The method
of RG cycling used in Ref.\cite{SU2_DENS} to study SU(2) gauge theory is one
of the best theoretically supported smoothing algorithms, but it is too
expensive in terms of both processor time and memory to be of any use
for SU(3) gauge theory.  In Ref.\cite{RG_SU2} RG cycling was fitted to a
series of APE smearing steps.  It was found that two APE steps with a
staple weight of 0.45 were equivalent to one RG cycling step for SU(2).
This method of smoothing the vacuum fluctuations will be referred to as
RG mapping.  RG mapping eliminates the expensive minimization step from
RG cycling by fitting one RG cycling step to a series of APE-smearing
steps.  

Since APE-smearing slowly distorts the topological content of the
lattice an extrapolation back to zero smearing steps is required.  This
means that the topological charge density must be measured at regular
intervals as one APE smears the lattice.  We found that the exact
parameters for APE smearing are not important as long as one monitors the
topological content over several steps and extrapolates back to zero
steps.

Generalizing the RG mapping method described above, we study the
instanton content of the SU(3) vacuum.  We use the same parameters for
the APE-smearing that were used for SU(2), a staple weight of 0.45 and
measurements of the topological density every two smearing steps.
Information before 12 smearing steps is discarded since the vacuum
fluctuations are still too high to reliably identify the instantons.  We
ran on pure gauge configurations with the Wilson action at couplings
$\beta = 5.85,6.0,6.1$.  We used two different lattice sizes, $12^4$
for $\beta = 5.85$ \& $6.0$, and $16^4$ for $\beta = $6.0 \& $6.1$.  A
detailed discussion of the method and results can found in
Ref.\cite{SU3_INST}.

\begin{figure}
\centerline{\ewxy{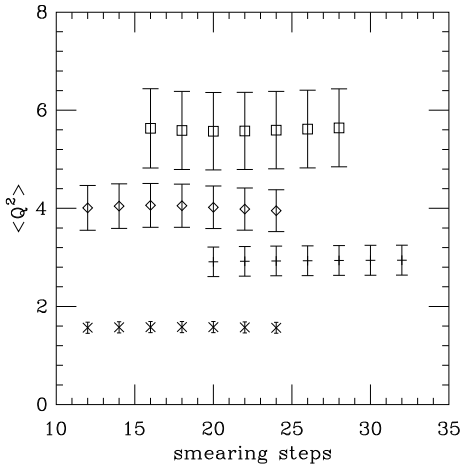}{80mm}}
\caption{The susceptibility $\langle Q^2\rangle$ vs number of
$c=0.45$ APE  steps.  Symbols are diamonds for $\beta=5.85$, crosses for
$\beta=6.0$ on $12^4$ lattices, squares for $\beta=6.0$ on $16^4$
lattices, and plusses for $\beta=6.1$.}
\label{fig:chi_vs_sweep}
% FIGURE_FILE Topo.ax
\end{figure}

The topological charge density was measured with a fixed point operator
every two smearing steps between approximately 15 and 30 steps depending
on the value of the coupling.  The total charge, $Q_{FP}$ gave an
integer value with 2-3 percent after 12 smearing steps.  The topological
susceptibility was calculated for the range of smearing steps for each
coupling value.  The susceptibility was very stable over smearing and
therefore there was no need to extrapolate back to zero smearing steps.
In Fig. \ref{fig:chi_vs_sweep} $<Q^2>$ is plotted against the number of
smearing steps done.  We arrive at a final value for the topological
susceptibility of $\chi^{1/4} = 203(5) MeV$.  Before one compares this
number with other works, we should note that we used a string tension
that is about 5\% higher than the standard value. If we had used the
more customary value of $\sqrt{\sigma} = 440MeV$ we would obtain a value
of $\chi^{1/4} = 192 MeV$, which is in complete agreement with the
results from Refs.\cite{Deforcrand_SU3,Teper_98,SCRI_susc}.

\begin{figure}
\centerline{\ewxy{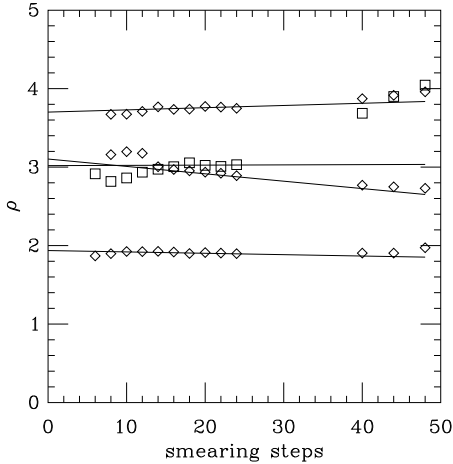}{80mm}}
\caption{Radius versus APE-smearing steps of instantons (for clarity, 
two symbols, diamonds and squares both denote instantons) and
anti-instantons (crosses) on a $16^4$ $\beta =6.0$ configuration.}
\label{fig:inst_evol}
% FIGURE_FILE inst.ax
\end{figure}

The instantons must be monitored since they are slowly distorted as the
lattice is smeared.  In in Fig. \ref{fig:inst_evol} we show the stable
objects for a typical lattice configuration.  The configuration has
topological charge $Q_{FP}$ that is within five percent of the integer
value of 4 from six smearing steps to 40 smearing steps.  Five stable
objects are found on the configuration, one anti-instanton and four
instantons.  The solid lines are the extrapolation to zero smearing
based on smearing steps 16 to 24. The slopes for all the extrapolations,
except the anti-instanton are less than 0.03.  Since that anti-instanton
size changes so rapidly and the extrapolation does not match the
evolution out to large smearing steps, we interpret this as a
misidentified vacuum fluctuation.  That conclusion is also supported by
the value of the total charge, $Q = 4$, on this configuration.  On a
typical configuration there are many ``lumps'' in the topological charge
density that could be naively identified as instantons.  Most of them
are not stable but disappear after one or a few smearing steps.  Further
support for this can be seen in the instanton density at various levels
of smearing.  After 12 smearing steps the density is $4.1 fm^{-4}$ and
after 24 smearing steps the density is $2.9 fm^{-4}$.  This is still
much higher than the phenomenologically expected value.  It is important
to separate the spurious objects from the true topological objects.
Our criterion is to keep only stable objects. An object is stable if it
can be identified at every analyzed smearing step and its size changes
slowly. Since the size of instantons change almost linearly with the
smoothing steps, we introduce a cut-off for the maximum acceptable
change.  This cut-off is chosen such that the topological susceptibility
calculated from the reliably measured total charge $Q_{FP}$ agrees with
that calculated from the charge $Q = I - A$ where I and A are the number
of stable instantons and anti-instantons on a given configuration.  When
applying this cut-off, the resulting instanton density turns out to be
$1 fm^{-4}$, in good agreement with the phenomenological expected value.

\begin{figure}
\centerline{\ewxy{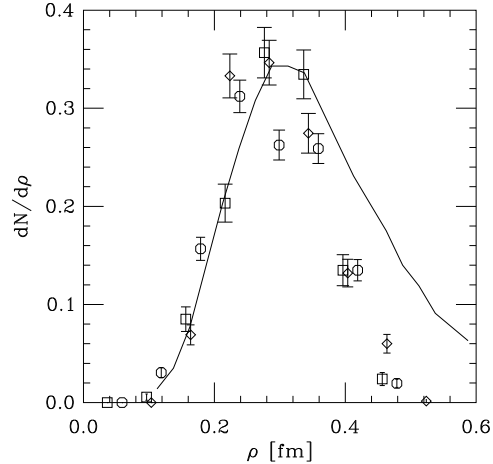}{80mm}}
\caption{The size distribution of the instantons.  The diamonds
correspond to $\beta=5.85$, octagons to $\beta=6.0$, and squares to
$\beta=6.1$.  The first bin of each distribution is contaminated by the
cut-off.  }
\label{fig:inst_dist}
% FIGURE_FILE dist.ax
\end{figure}

In Fig. \ref{fig:inst_dist} we show our instanton size distribution.
The solid line in Fig. \ref{fig:inst_dist} is the instanton size
distribution from the instanton liquid model for $\Lambda_{MS} =
200MeV$, provided by Shuryak.  Our distribution along with our values
for density and the average instanton size put us in good agreement with
the various phenomenologically successful instanton liquid models.

We can now give a final description of the RG mapping method.  RG
mapping is a series of APE-smearing steps where the topological content
is monitored at regular intervals over a range of smearing steps.  The
original content of the lattice is then found by making a linear
extrapolation to zero smearing steps while only accepting objects whose
extrapolation has slope below a certain cut off.  This cut-off is tuned
so that $<Q_{FP}^2> = <(I-A)^2>$ where Q is the topological charge measured
with the fixed point operator and I and A are the number instantons and
anti-instantons found.

Using this method we calculate the relevant parameters of the instanton
vacuum for SU(3).  We find an instanton density of $1 fm^{-4}$ and an
average instanton size of approximately $0.3 fm$.  This is considerable
smaller than the results given in Ref.\cite{Deforcrand_SU3,Teper_98}.
We attribute this difference to our extrapolation to zero smearing steps.

\end{document}